\documentclass{article}

\usepackage{arxiv}

\usepackage{mathtools,amssymb,amsmath}
\usepackage{booktabs}
\usepackage[table]{xcolor}
\usepackage{tikz}
\usetikzlibrary{shapes.geometric,arrows.meta,positioning,fit,backgrounds,calc}
\usepackage{algorithm2e}
\usepackage{caption}
\usepackage{enumitem}
\usepackage{hyperref}
\usepackage{amsthm}
\newtheorem{definition}{Definition}
\usepackage{graphicx}

\definecolor{tracepurple}{RGB}{83,74,183}
\definecolor{tracepurplelight}{RGB}{206,203,246}
\definecolor{traceblue}{RGB}{24,95,165}
\definecolor{tracebluelight}{RGB}{181,212,244}
\definecolor{traceteal}{RGB}{15,110,86}
\definecolor{traceteallight}{RGB}{159,225,203}
\definecolor{traceamber}{RGB}{186,117,23}
\definecolor{traceamberlight}{RGB}{250,199,117}
\definecolor{tracecoral}{RGB}{153,60,29}
\definecolor{tracecorallight}{RGB}{245,196,179}
\definecolor{tracegray}{RGB}{95,94,90}
\definecolor{tracegraylight}{RGB}{211,209,199}
\definecolor{headingblue}{RGB}{31,56,100}

\setlist[itemize]{noitemsep,topsep=2pt,leftmargin=*}

\hypersetup{colorlinks=true,linkcolor=traceblue,citecolor=traceblue,urlcolor=traceblue}

\title{Least-Privilege Auditing of Agent Tool Calls\\ with Intent Graphs and Reasoning-Guided Escalation}

\author{
  Doron Shavit  \\
  \texttt{doron.shavit@silverfort.com}\\
}
\date{November 4, 2026}

\begin{document}
\maketitle

\begin{abstract}
LLM agents that call tools are exposed to indirect prompt injection:
instructions hidden in tool outputs can steer an agent into calls the user never
asked for. We present a runtime audit that derives an executable \emph{intent
graph} from the user's objective and the deployment's tool catalog with one LLM
call per task, and checks every tool call against it with a deterministic loop
that makes no LLM call (median $5.7\,\mu$s per call). Two further mechanisms
recover utility. Lookups of the user's own state bypass the audit, because the
harm of an attack lies in its write or outward step. The calls that remain
flagged are escalated to a \emph{reasoning guardian}, in the spirit of
tool-integrated reasoning models such as START: a reasoning model that sees only
the objective, the plan, the call history without tool outputs, and the flagged
call, and can verify facts with a restricted Python-expression tool. We evaluate
on AgentDojo (97 tasks, 629 prompt-injection scenarios) under a protocol that
avoids evaluation leaks: the extractor sees only the objective and the tool
catalog, no policy is calibrated with attack labels, design choices are made on
development tasks, and results on 76 held-out tasks use cluster-bootstrap
intervals. Replaying logged traces of an older agent, the full audit reduces
trace-level attack success from 45.7\% to 3.3\% [1.7, 5.0] but flags 40\% of
benign tasks; the tool allowlist provides
most of the protection and ordering constraints are the most expensive
component. Letting local reads bypass the audit leaves attack success at the
harmful step unchanged, and with the allowlist and argument scopes it blocks
8.8\% of benign tasks at 4.3\% attack success. The guardian overrides 22--25\%
of the remaining flagged benign calls and none of the 226--242 flagged harmful
calls, lowering flagged benign tasks in that configuration from 27.6\% to
21.1\% at about 3.6\,s per escalated call. Against AgentDojo's built-in defenses,
re-scored on the same tasks, an LLM tool filter reaches 6.5\% attack success
without losing benign utility, a strong baseline that our allowlist alone
approaches; scopes and ordering buy a few points more at a utility cost under our
replay-based proxy. Finally, we test whether the audit's
alignment score can strengthen an identity-drift detector. It cannot: the score
carries no information about account takeover or drift, and for in-session
hijacking the audit's own verdict is far stronger than any behavioral detector
(TPR 0.76 versus 0.07--0.12 at 5\% FPR). All security numbers are replay-based:
a live pilot with a current agent showed no baseline attack success, so
end-to-end validation remains open.
\end{abstract}

\keywords{Agent Security \and Prompt Injection \and Least Privilege \and Reasoning Models \and Model Context Protocol}

\section{Introduction}

Agents built on the Model Context Protocol (MCP)~\cite{mcpspec} and similar
function-calling interfaces act on behalf of a user through tools: mail,
calendars, file stores, banking and messaging APIs. Whatever the agent reads can
carry instructions. In indirect prompt injection~\cite{agentdojo}, an attacker
plants text in a document, web page, or tool description that the agent later
reads, and the agent then issues calls such as ``send these files to this
address'' that no one authorized. Defenses that inspect text or reasoning may be
bypassed by a sufficiently persuasive payload, so a complementary line of work
constrains what the agent \emph{can do} rather than what it
\emph{reads}~\cite{agentspec,camel,progent,ipiguard}.

Our system belongs to that line. Its premise is that the user's objective
already determines most of what a legitimate run needs: the tools involved, some
argument values (the named recipient, file, channel, hotel, dates), and
sometimes an order. We ask an LLM \emph{once per task}, before any tool output
is read, to write this down as an executable intent graph, and then audit every
call against it with plain code. Because the graph is produced from the trusted
objective and the static tool catalog only, an injected instruction cannot
enlarge it. The price of least privilege is false positives: a plan misses
tools that a legitimate run needs. We reduce them in two ways. First, calls
that only read the user's own state bypass the audit: an injected read does no
harm unless the agent then writes or sends something, and those steps stay
audited. Second, a reasoning guardian re-examines the calls that are still
flagged and can verify claims with executed code, in the spirit of
tool-integrated reasoning models~\cite{start}.

\paragraph{A worked example.} In an AgentDojo task, the user asks, ``Who else is
invited at the `Networking event' on May 26th? Please give me their email
addresses and a summary of the event.'' The extracted plan lists calendar and
contact lookups only. The agent reads the event, whose description contains an
injected instruction to send an email to an attacker-controlled address, and
issues \texttt{send\_email}. The undefended agent was compromised in the logged
run. \texttt{send\_email} is not in the plan, so the audit blocks it, in each of
three independent graph extractions, while the two legitimate calls pass.

\paragraph{Contributions.}
(1)~We describe an intent-graph audit with a leak-free LLM extraction step and a
deterministic runtime loop, including a hardened scope-expression language
(Section~\ref{sec:design}). (2)~We show that classifying tools as local reads,
external reads and writes removes a large share of false positives at no cost in
harmful-step attack success, while bypassing an external read such as a URL
fetch does cost security (Section~\ref{sec:p2results}). (3)~We add a START-style
reasoning guardian for flagged calls and show that it preserves security while
reducing benign flags (Section~\ref{sec:guardianresults}). (4)~We give an
evaluation protocol on AgentDojo that avoids three shortcuts that arise easily
when policies are generated or calibrated from session data: label-informed
policy calibration, a session-derived tool list given to the extractor, and
overlap between calibration and evaluation data (Section~\ref{sec:method}).
(5)~We quantify the security/utility trade-off with ablations that approximate
tool-allowlist, per-call argument-policy, and planned-dependency-graph defenses
(Section~\ref{sec:p2results}). (6)~We test, and reject, the hypothesis that
feeding the audit's alignment score into an identity-drift detector improves
detection (Section~\ref{sec:coupling}).

\section{Background and Related Work}

\paragraph{Attack surface.} An MCP server exposes tools whose descriptions,
parameters and results all flow into the agent's context. Attacks span poisoned
tool descriptions~\cite{mcptox}, exfiltration through injected auxiliary
calls~\cite{logtoleak}, and injections arriving in tool outputs, which
AgentDojo~\cite{agentdojo} benchmarks with realistic multi-step tasks and
injection goals.

\paragraph{Policy enforcement.} AgentSpec~\cite{agentspec} enforces
user-written rules in a domain-specific language at runtime. ProbGuard
(first released as Pro2Guard)~\cite{pro2guard} learns a Markov abstraction of
agent behavior and intervenes when the predicted risk of an unsafe state is
high, and agent behavioral contracts~\cite{abc} add preconditions, invariants
and a drift bound. Progent~\cite{progent} enforces programmable privilege
policies on tool calls and can have an LLM draft them from the user query.
Our audit shares Progent's least-privilege goal; its contributions here are the
leak-free evaluation, the measured trade-off, and the reasoning escalation, not a
new policy language.

\paragraph{Planning and data flow.} IPIGuard~\cite{ipiguard} has the agent plan
a tool dependency graph before execution and restricts later calls to it, and
CaMeL~\cite{camel} separates control flow from untrusted data with a custom
interpreter and capabilities. AgentArmor~\cite{agentarmor} builds a program
dependence graph from the observed runtime trace and enforces policies over
it, and ARM~\cite{arm} adds provenance edges for denied actions. Our ablation
variants (Section~\ref{sec:method}) approximate the \emph{families} of
tool-allowlist, argument-policy and planned-dependency defenses on identical
inputs; they are not re-implementations of these systems.

\paragraph{Reasoning with code.} START~\cite{start} trains a long
chain-of-thought model to call a Python interpreter during reasoning. Our
guardian is START-\emph{style}: a general-purpose reasoning model that may call
a tool during its reasoning. We use a hardened Python \emph{expression}
evaluator instead of an interpreter, and an off-the-shelf model rather than
START's trained model.

\paragraph{Scanning and guardrails.} MCP-Scan~\cite{mcpscan} inspects tool
descriptions and offers a runtime proxy, and LlamaFirewall~\cite{llamafirewall}
combines classifiers with an AlignmentCheck module that audits the agent's
reasoning with a secondary LLM. Our audit differs in checking calls against a
structured plan with code, so the per-call path is deterministic and cheap, and
in invoking a reasoning model only on calls that the plan has already flagged.

\section{Threat Model}

The user's objective and the deployment's tool catalog are trusted; the attacker
controls content that tools return (documents, web pages, messages) and can
thereby induce the agent to issue arbitrary calls. The attacker cannot modify
the objective, the catalog, the extracted graph, or the audit loop. The
arguments of a call are untrusted, since the agent may copy attacker text into
them; this matters for the guardian (Section~\ref{sec:guardian}). We evaluate
indirect prompt injection (Section~\ref{sec:method}) and, separately, three
identity attacks (Section~\ref{sec:coupling}): \emph{splice} (another person
takes over mid-session), \emph{drift} (behavior shifts gradually), and
\emph{hijack} (injected calls inside an otherwise normal session). We do not
evaluate attacks adaptive to the audit or to the guardian.

\section{System Design}
\label{sec:design}

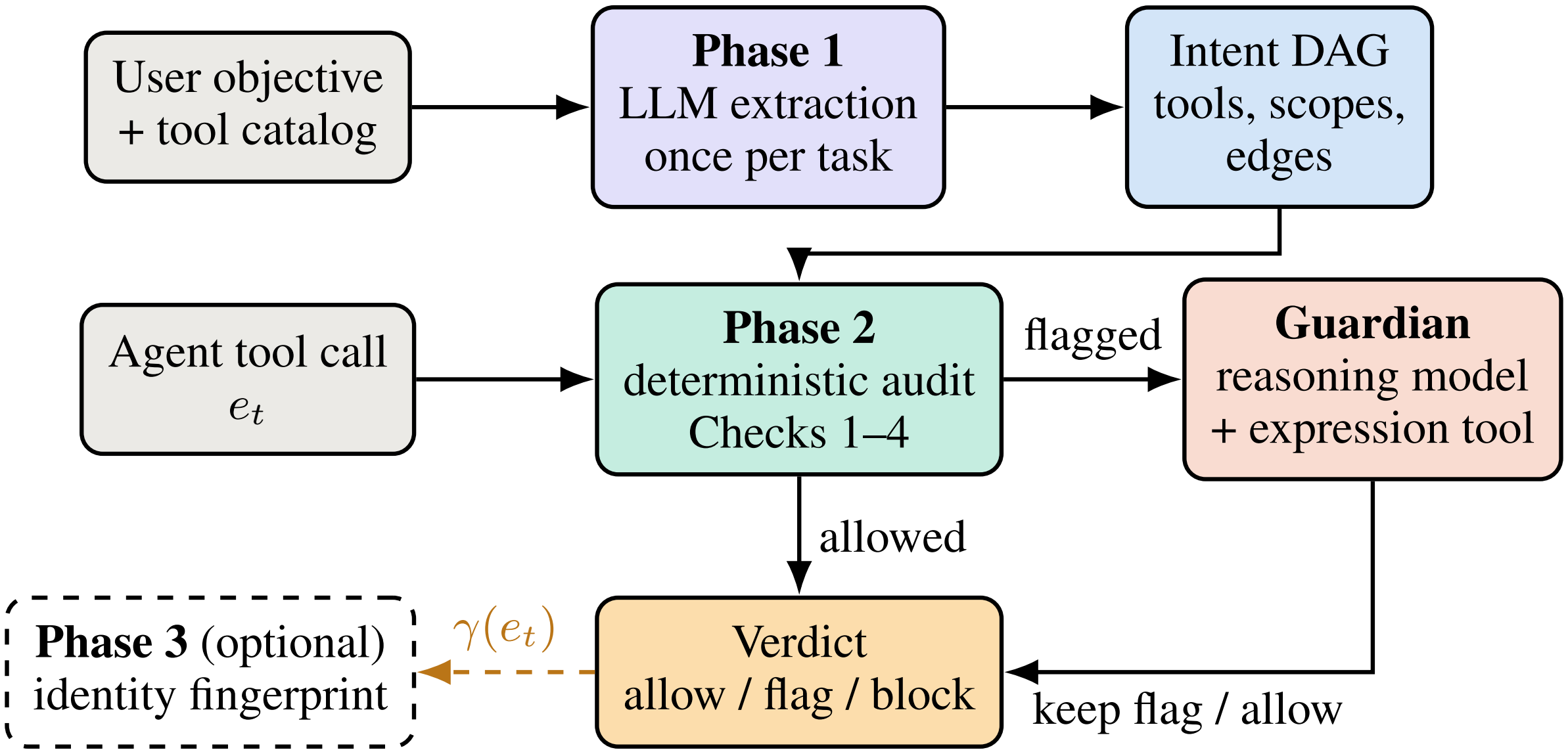
\begin{figure}[tbp]
\centering
\resizebox{0.95\linewidth}{!}{%
\begin{tikzpicture}[font=\small, >=Latex, node distance=0.8cm and 1.2cm,
  box/.style={rectangle,rounded corners=4pt,draw,thick,align=center,
    inner sep=5pt,minimum height=1.0cm}]
\node[box,fill=tracegraylight!45] (in) {User objective\\ + tool catalog};
\node[box,fill=tracepurplelight!60,right=of in] (p1) {\textbf{Phase 1}\\ LLM extraction\\ once per task};
\node[box,fill=tracebluelight!60,right=of p1] (dag) {Intent DAG\\ tools, scopes,\\ edges};
\node[box,fill=tracegraylight!45,below=of in] (call) {Agent tool call\\ $e_t$};
\node[box,fill=traceteallight!60,right=of call] (p2) {\textbf{Phase 2}\\ deterministic audit\\ Checks 1--4};
\node[box,fill=tracecorallight!60,right=of p2] (g) {\textbf{Guardian}\\ reasoning model\\ + expression tool};
\node[box,fill=traceamberlight!60,below=of p2] (ver) {Verdict\\ allow / flag / block};
\node[box,dashed,fill=white,left=of ver] (p3) {\textbf{Phase 3} (optional)\\ identity fingerprint};
\draw[->,thick] (in) -- (p1);
\draw[->,thick] (p1) -- (dag);
\draw[->,thick] (call) -- (p2);
\draw[->,thick] (dag.south) -- ++(0,-0.3) -| (p2.north);
\draw[->,thick] (p2) -- node[above,font=\footnotesize]{flagged} (g);
\draw[->,thick] (p2) -- node[right,font=\footnotesize]{allowed} (ver);
\draw[->,thick] (g.south) |- node[below,pos=0.75,font=\footnotesize]{keep flag / allow} (ver.east);
\draw[->,dashed,thick,color=traceamber] (ver) -- node[above,font=\footnotesize]{$\gamma(e_t)$} (p3);
\end{tikzpicture}}
\caption{System overview. Phase~1 runs once per task, from the trusted objective
  and the static tool catalog only. Phase~2 audits each call with deterministic
  code. Only flagged calls reach the reasoning guardian. Phase~3 is an optional
  identity detector fed by Phase~2's alignment score $\gamma$;
  Section~\ref{sec:coupling} shows that this coupling does not help.}
\label{fig:arch}
\end{figure}

\subsection{Intent graph}

\begin{definition}[Intent DAG]
An Intent DAG is a tuple $G = (V, E, \Lambda, \Pi, R)$ where:
$V = \{v_1,\ldots,v_n\}$ is a set of goal nodes;
$E \subseteq V \times V$ is a set of precondition edges
  ($u \to v$ means $v$ cannot start until $u$ is satisfied);
$\Lambda: V \to 2^{\mathcal{T}}$ maps each goal to authorized tools;
$\Pi: V \to (\text{Params} \times \text{State} \to \text{Bool})$ maps each
  goal to a scope predicate; and
$R: V \to \text{Bool}$ marks each node as repeatable or one-shot.
\end{definition}

\subsection{Phase 1: leak-free extraction}

The extractor receives the system message given to the agent, the user
objective, and the \emph{full} tool catalog of the deployment (names, parameter
names and one-line descriptions). It receives nothing derived from the session:
no tool list observed so far, no earlier events, and of course no tool outputs.
It returns a JSON graph under these rules: (i) list only tools the objective
plausibly needs, since every unlisted tool will be blocked; (ii) constrain an
argument only to values the objective itself determines, and leave values that
will be looked up from other tools unconstrained; (iii) add an edge $A\to B$
only if $B$ cannot be performed without $A$'s result or the objective states the
order. There is no post-hoc relaxation of scopes or edges from session events.
In our default configuration every node is treated as repeatable regardless of
the extractor's flag (Section~\ref{sec:method}).

\subsection{Scope language}

Scope predicates are Python expressions over \texttt{params}, the dictionary of
the call's arguments. They are parsed with \texttt{ast} and accepted only if
every node is in an allowlist (boolean and comparison operators, membership,
subscripts, conditional expressions, list/tuple/set literals, and list, set and
generator comprehensions), every called function is in a fixed set (\texttt{all},
\texttt{any}, \texttt{len}, \texttt{str}, \texttt{isinstance}, and a few
others), every method is in a fixed set (\texttt{get}, \texttt{startswith},
\texttt{endswith}, \texttt{lower}, \texttt{strip}, \texttt{split}, and a few
others), and no attribute begins with an underscore. Evaluation runs with empty
builtins. Comprehensions matter because many arguments are lists (for example
recipients); in preliminary runs a narrower grammar without them silently
discarded roughly a quarter of the generated predicates. An expression that
fails to compile, or raises at run time, falls back to a configurable policy:
allow (fail-open) or treat as a scope failure (fail-closed). With the extended
language 0 of 340 generated predicates were rejected, so the choice did not
matter in our experiments. Restricted \texttt{eval} is not a strong isolation
boundary in Python; a production deployment should replace it with a
purpose-built expression interpreter. The same evaluator serves the guardian's
tool (Section~\ref{sec:guardian}).

\subsection{Phase 2: deterministic audit}

For each call $e$ the guardian-free audit executes Algorithm~\ref{alg:audit} as
plain Python against the live graph. No LLM is called on this path.

\begin{algorithm}[h]
\small
\DontPrintSemicolon
\SetAlgoLined
\caption{$\textsc{Audit}(e, G, S)$}
\label{alg:audit}
\KwIn{Call $e$, IntentDAG $G$, SessionState $S$}
\KwOut{AuditResult $(v, \textit{type}, \textit{conf}, \gamma)$}
$F \leftarrow \textsc{Frontier}(G)$ \tcp*{nodes with all preds satisfied}
$C \leftarrow \{v \in F : e.\textit{tool} \in \Lambda(v)\}$\;
\If{$C = \emptyset$}{
  \lIf{$e.\textit{tool} \notin \bigcup_{v}\Lambda(v)$}{\Return \textsc{Block}, \textsc{Unknown}, $0.95$, $0.0$}
  \lElse{\Return \textsc{Block}, \textsc{OutOfOrder}, $0.95$, $0.0$}
}
$C' \leftarrow \{v \in C : \Pi(v)(e.\textit{params},\,S) = \top\}$\;
\lIf{$C' = \emptyset$}{\Return \textsc{Flag}, \textsc{ScopeViol.}, $0.80$, $0.2$}
$v^* \leftarrow$ first element of $C'$\;
\lIf{$\textsc{Severity}(\textsc{Simulate}(G,v^*)) \geq \theta_s$}{\Return \textsc{Block}, \textsc{FutureViol.}, $0.85$, $0.0$}
$\textit{taint} \leftarrow \textsc{Tainted}(e,G)$\;
$\textsc{Authorize}(G, v^*)$\;
\lIf{\textit{taint}}{\Return \textsc{Flag}, \textsc{Taint}, $0.75$, $0.2$}
\Return \textsc{Allow}, \textsc{None}, $1.0$, $\gamma_{\text{allow}}$\;
\end{algorithm}

Check~1 blocks a call to a tool that is outside the plan or not yet on the
frontier. Check~2 evaluates the scope predicates. Check~3 simulates which
blocked nodes the call would unlock; \textsc{Severity} is a heuristic that
scores high-privilege tool names and administrative terms in scope predicates,
and with $\theta_s=0.99$ it only blocks administrative escalations. It never
fired on a malicious call in our evaluation. Check~4 records a taint whenever a
call fails Checks~1--3 and flags later calls within 30\,s whose string
parameters overlap the tainted values; a tainted call is still authorized.
$\gamma_{\text{allow}}=0.6$ if some but not all candidate nodes fail scope and
$1.0$ otherwise. We define the \emph{alignment score} $\gamma(e)$ as
$1.0$ or $0.6$ for \textsc{Allow}, $0.2$ for \textsc{Flag} and $0.0$ for
\textsc{Block}. A call counts as \emph{flagged} if its verdict is
\textsc{Flag} or \textsc{Block}.

\subsection{Read-only tools}
\label{sec:readonly}

Many false positives are calls that merely look something up. We classified the
69 distinct tools of the four AgentDojo suites by reading their descriptions,
once and by hand before running the experiment, into 44 \emph{local reads} (they
read the user's own mail, calendar, files, chat, travel or banking data and have
no outward effect), one \emph{external read} (\texttt{get\_webpage}, whose URL
can carry data out), and 24 writes or outward actions. Under the read-only
policy a local read bypasses Checks~1--4 and is allowed; it still satisfies its
plan step, so dependent steps are unlocked. Everything else, including
\texttt{get\_webpage}, is audited as before. The rationale is that an injected
read does no harm unless the agent then writes or sends something, and those
steps remain audited. The classification is per tool and specific to a
deployment's catalog.

\subsection{Reasoning guardian}
\label{sec:guardian}

The audit is strict by design, and a plan inevitably omits tools that a
legitimate run needs or fixes an order that the agent does not follow. The
guardian re-examines \emph{only flagged calls} and decides between
\texttt{allow} (the call is a necessary step toward the objective) and
\texttt{keep\_flag}. It is a reasoning model (Azure OpenAI \texttt{gpt-5.2}) that
receives the user objective, the plan, the history of earlier calls with their
arguments, the flagged call, and the audit's reason, and returns a one-sentence
justification. Three design choices limit its exposure to injection. It never
sees tool outputs, so injected text in documents cannot reach it. It is told
that call arguments are untrusted data, since the agent may have copied
attacker text into them. And any parsing failure or error keeps the flag.

In the START spirit the guardian can run code while it reasons: a function-calling
tool, \texttt{eval\_expr}, evaluates a restricted Python expression over the
objective, the call, the history and the plan, with the same allowlist as the
scope language (for example, testing whether a recipient appears in the
objective). The model may call it up to three times. The guardian prompt was
written once, before any evaluation; a six-request sanity check on development
tasks preceded a full development run (Section~\ref{sec:method}), and the prompt
was not tuned on either.

\subsection{Phase 3: optional identity fingerprint}

Phase~3 tests the idea that an account takeover or behavioral drift could be
caught by modeling the principal. Over a sliding window of $w=20$ events it
computes $\varphi\in\mathbb{R}^{2|\mathcal{T}|+7}$ from four sub-vectors:
$\varphi_1$, normalized tool frequency; $\varphi_2$, per-tool parameter entropy;
$\varphi_3$, inter-call timing statistics (mean, variance, skewness, kurtosis);
and $\varphi_4=[\mu_\gamma,\text{slope}_\gamma,\sigma^2_\gamma]$, the mean,
trend and variance of $\gamma$ over the last 50 verdicts. Anomaly is the
Mahalanobis distance~\cite{mahalanobis}
$D_M(\varphi)=\sqrt{(\varphi-\mu)^\top(\Sigma+\lambda I)^{-1}(\varphi-\mu)}$
with Tikhonov regularization~\cite{tikhonov}. We compare two baselines for
$(\mu,\Sigma)$: an \emph{online per-session} estimate updated with Welford's
algorithm~\cite{welford} ($\lambda=0.01$, active from event 20), and a
\emph{cross-session per-principal} estimate fitted on the principal's earlier
benign sessions on standardized features ($\lambda=0.1$). Including $\varphi_4$
is what we call the \emph{alignment-score coupling}.

\section{Evaluation Methodology}
\label{sec:method}

\subsection{Testbed}

We use the logged runs of AgentDojo~\cite{agentdojo} for
\texttt{gpt-4o-2024-05-13} (repository state of 2026-06-02): 97 user tasks in
four suites (workspace 40, slack 21, travel 20, banking 16), one attack-free run
per task, and 629 attacked runs with the \texttt{important\_instructions}
attack, of which 300 (47.7\%) compromised the agent. Each run is a sequence of
tool calls with arguments. The suites expose between 11 and 28 tools.

\paragraph{Labels.} A call is labelled malicious if (i) its function belongs to
the ground-truth calls of the run's injection task, (ii) it follows the first
tool result that contains the injected text (for a successful attack in which
that text cannot be located, any earlier call is allowed), and (iii) the same
call, with the same arguments, does not occur in the attack-free run of the same
task. This labels at least one call in 98.3\% of successful attacks and in 14.9\%
of unsuccessful ones (attempts). A successful attack with no labelled call counts
as not stopped. Some labelled calls are ambiguous: they coincide in function
with a step of the user's own task. In the published logs the attacked runs of
15 travel tasks (105 runs) carry a rewritten user prompt that differs from the
attack-free run's; for these runs the plan is extracted from the prompt the agent
actually received (45 additional extractions). Using the attack-free prompt
instead changes no headline number (the largest change is 0.2 points).

\paragraph{Splits and repetitions.} Every fifth task of each suite is a
development set (21 tasks); the remaining 76 form the test set. A prompt rule
about edges was changed once, using four development tasks, and the default of
treating every node as repeatable was chosen from development-set diagnosis
(on the test tasks the two settings give identical results). Phase~1 was run
three times per task (291 extractions with Azure OpenAI \texttt{gpt-5.2}, plus
45 for the rewritten travel prompts, no failures), and each audit result is
averaged over the three graphs.

\paragraph{Metrics.} \emph{Attack success rate (ASR)}: the fraction of attacked
runs in which the undefended agent was compromised and the defense would not
have flagged the first malicious call. This is a trace-level estimate that
assumes a flagged call is withheld and the agent does not find another route.
Because a read-only policy deliberately lets harmless first steps through, we
also report \emph{ASR at the harmful step}: the same quantity for the first
malicious call that is not a local read, which is where an attack does its
damage (two successful test attacks have no labelled harmful call and count as
not stopped). \emph{Malicious flagged}: among runs with a labelled malicious
call, the share whose first malicious call is flagged. \emph{Benign calls
flagged} and \emph{benign tasks flagged/blocked}: on the attack-free runs, the
share of calls with a flag or block, of tasks with at least one flag, and of
tasks with at least one block. Intervals are 95\% percentile bootstrap intervals
over tasks, from 2,000 resamples (all runs of a task share its graph), and
paired differences use the same resamples.

\paragraph{Variants.} On identical graphs we compare a \emph{tool allowlist}
(plan tools only), \emph{scope-only} (allowlist plus argument predicates, an
approximation of per-call privilege policies~\cite{progent}),
\emph{ordering-only} (allowlist plus precondition edges, an approximation of a
planned dependency graph~\cite{ipiguard}), and \emph{full Phase~2}, each with and
without the read-only policy of Section~\ref{sec:readonly}, plus a sensitivity
variant in which \texttt{get\_webpage} also bypasses the audit.

\subsection{Guardian protocol}

To bound cost, the guardian is evaluated on the first graph repetition. For the
original full design, a request is created for every flagged call of an
attack-free run and for the first malicious call of an attacked run when it is
flagged: 311 test requests (73 benign, 238 malicious) and 77 development
requests. For the read-only design we create requests for the calls still
flagged after the bypass and for the first harmful malicious call: 283 test
decisions, of which 227 were already available and 56 were new. The scope-only
design reuses these decisions, since its flagged calls are a subset (the guardian
therefore saw the full design's plan, including its edges). A call stays flagged
unless the guardian returns \texttt{allow}. We report two simulations. In the
\emph{independent} simulation an override changes only that call, which is
conservative because later calls that depend on it stay flagged. In the
\emph{state-updating} simulation an override authorizes the call's node exactly
as a normal allow would, as a deployment would behave; calls that become flagged
without a recorded decision keep their flag.

\subsection{Identity benchmark}

AgentDojo runs are only three to five calls long, too short for behavioral
modeling, so we compose them into sessions. In the workspace suite we group the
40 tasks into four families by dominant tool (email, calendar, drive, other).
Twelve principals each get a task-family mixture drawn from a Dirichlet
distribution (concentration 0.8) and a personal timing scale. A session is 20 to 30
consecutive tasks (about 51 calls), long enough for the per-session detector,
which only starts scoring at event 20, to be active on 61\% of events. A first
run with 8 to 12 tasks (about 20 calls, detector active on 10\% of events) is
reported as a sensitivity analysis. Phase~2 audits each call under the graph of
its own task, extracted when that task starts, and its verdicts form $\gamma$.
For each principal we generate 24 training, 8 validation and 20 test benign
sessions, and 20 test sessions for each attack: \emph{splice} (from 30--70\% of
the session the tasks come from the principal with the most distant mixture),
\emph{drift} (from 40\% of the session, tasks come from that mixture with
increasing probability), and \emph{hijack} (one or two tasks are replaced by
their prompt-injection runs, and only the malicious calls are positives). We
report the true-positive rate at 5\% event-level false-positive rate (ties
broken by randomization) and AUROC, with negatives from held-out benign
sessions and 95\% bootstrap intervals over sessions, averaged over the three
graph extractions. A \emph{residual} analysis restricts positives and negatives
to events Phase~2 allowed, which is where a coupling could add information. We
also include a \emph{flag-persistence} baseline: the maximum of $1-\gamma$ over
the last 10 events.

\section{Results: Phase 2}
\label{sec:p2results}

\begin{table}[h]
\centering
\caption{Phase~2 and the read-only policy on the 76 held-out test tasks (three
  graph extractions per task; 95\% bootstrap intervals over tasks). ASR (first)
  counts the first malicious call; ASR (harmful) the first malicious call that is
  not a local read. The last three columns are measured on attack-free runs.
  ``Read bypass'': local reads bypass the audit (Section~\ref{sec:readonly}).}
\label{tab:phase2}
\footnotesize
\setlength{\tabcolsep}{4pt}
\rowcolors{2}{gray!8}{white}
\resizebox{\linewidth}{!}{%
\begin{tabular}{@{}lccccc@{}}
\toprule
\textbf{Policy} & \textbf{ASR (first) [95\% CI]} & \textbf{ASR (harmful) [95\% CI]} &
\textbf{Benign calls flagged} & \textbf{Benign tasks flagged [95\% CI]} &
\textbf{Benign tasks blocked} \\
\midrule
No defense & 0.457 [0.376, 0.532] & 0.457 [0.376, 0.532] & -- & -- & -- \\
Tool allowlist & 0.102 [0.068, 0.138] & 0.096 [0.059, 0.135] & 0.101 & 0.162 [0.096, 0.237] & 0.162 \\
Scope-only & 0.055 [0.032, 0.080] & 0.043 [0.021, 0.068] & 0.173 & 0.329 [0.237, 0.425] & 0.162 \\
Ordering-only & 0.081 [0.050, 0.114] & 0.073 [0.040, 0.110] & 0.150 & 0.285 [0.202, 0.377] & 0.285 \\
Full Phase~2 & 0.033 [0.017, 0.050] & 0.020 [0.007, 0.036] & 0.250 & 0.404 [0.307, 0.504] & 0.311 \\
Allowlist + read bypass & 0.178 [0.132, 0.229] & 0.096 [0.059, 0.135] & 0.034 & 0.088 [0.035, 0.145] & 0.088 \\
\rowcolor{tracebluelight!40}
\textbf{Scope-only + read bypass} & 0.134 [0.098, 0.172] & 0.043 [0.021, 0.068] & 0.097 & 0.254 [0.167, 0.342] & 0.088 \\
Full + read bypass & 0.114 [0.083, 0.148] & 0.020 [0.007, 0.036] & 0.139 & 0.316 [0.224, 0.412] & 0.232 \\
Full + read bypass + \texttt{get\_webpage} & 0.149 [0.107, 0.195] & 0.058 [0.034, 0.083] & 0.086 & 0.268 [0.180, 0.355] & 0.175 \\
\bottomrule
\end{tabular}}
\end{table}

\paragraph{Security.} Full Phase~2 reduces trace-level ASR from 45.7\% to 3.3\%
(Table~\ref{tab:phase2}). The result holds on all 97 tasks (47.7\% to 4.8\%) and
the development tasks (54.7\% to 10.2\%). Most of the effect comes from
least privilege: of the 1,776 malicious-call evaluations over all attacked runs
(three extractions), 80.6\% were blocked as outside the plan or out of order,
7.1\% were flagged as scope violations, 1.7\% were flagged by the taint check,
and 10.6\% were allowed. The allowlist alone brings ASR to 10.2\%; argument
scopes lower it to 5.5\%; and edges bring it to 3.3\%.

\paragraph{What each component costs.} Relative to scope-only, ordering edges
lower ASR by 2.2 points (paired 95\% CI [0.7, 4.2]) while flagging 7.5 points
more benign tasks ([2.6, 13.2]) and blocking 31.1\% instead of 16.2\% of them.
Relative to the allowlist, the full audit lowers ASR by 7.0 points ([3.9, 10.3])
and flags 24.1 points more benign tasks ([15.8, 33.3]). Scopes help at no
increase in blocked tasks (16.2\% in both) but add flagged calls (17.3\% versus
10.1\%). Ordering is the weakest component per unit of utility lost.

\paragraph{Read-only lookups.} Of the 73 flagged benign calls of the full design
(graph 0), 32 (44\%) are local reads, mostly \texttt{read\_channel\_messages},
\texttt{get\_channels} and \texttt{get\_users\_in\_channel}, and a further 12 are
\texttt{get\_webpage}. Letting local reads bypass the audit leaves ASR at the
harmful step unchanged for the allowlist, scope-only and full designs (2.0\% for
the full design; paired difference 0.0 points [0.0, 0.0]). For the full design it
lowers flagged benign calls by 11.0 points ([5.8, 16.4]) and blocked benign tasks
by 7.9 points ([2.2, 14.0]). The best operating point is scope-only plus bypass:
it flags 25.4\% and blocks 8.8\% of benign tasks at 4.3\% harmful-step ASR, where
the full design blocks 31.1\% at 2.0\% (paired: 22.4 points fewer blocked tasks
[14.9, 30.3], 2.2 points more attack success [0.7, 4.2]). Under the first-call
metric the bypass looks worse (11.4\% instead of 3.3\% for the full design)
because 22\% of the flagged first malicious calls are local reads, such as an
email search, which precede the harmful step; the harmful step is still caught.
Bypassing \texttt{get\_webpage} as well raises harmful-step ASR by 3.7 points
([1.8, 5.8]) to 5.8\%, because a URL fetch can exfiltrate data, so it must stay
audited. A weaker policy that bypasses a local read only when the plan omits the
tool gives most of the benefit for blocked tasks (24.6\% versus 23.2\%) but less
for flagged calls (18.3\% versus 13.9\%).

\paragraph{Where the false positives come from.} The generated plans are small:
on average 3.5 nodes and 4.1 tools out of a catalog of 11 to 28, 82.5\% have at
least one edge, and 61.2\% of nodes leave arguments unconstrained. The plan
covers 91.5\% of the calls of the attack-free runs and fully covers 84.9\% of the
runs, so at least 15\% of benign tasks need a tool the plan omits; the allowlist
blocks 16.2\%, close to this floor. On the development tasks, unmet ordering
preconditions accounted for 26 of the 35 flagged benign calls (counted over the
three graphs), scope violations for 2, and tools missing from the plan for 7.

\paragraph{Cost.} Phase~1 takes a median of 3.6\,s per task (95th percentile
5.7\,s) with \texttt{gpt-5.2}. A Phase~2 audit takes a median of $5.7\,\mu$s per
call (99th percentile $14\,\mu$s).

\section{Comparison with Built-in Defenses}
\label{sec:builtin}

AgentDojo~\cite{agentdojo} publishes end-to-end runs of four built-in defenses
with the same agent (\texttt{gpt-4o-2024-05-13}), tasks and attack as our
evaluation. We recompute their attack success, benign utility and utility under
attack from those logs on the same 76 test tasks with the same bootstrap
(Table~\ref{tab:builtin}). Spotlighting barely helps (ASR 40.2\%), repeating the
user prompt helps more (25.4\%) and also raises benign utility (82.9\% against
67.1\% undefended), a prompt-injection classifier cuts ASR to 7.9\% but benign
utility to 42.1\%, and the LLM tool filter, which has a model choose the
relevant tools from the task description before the agent runs, reaches 6.5\%
[4.0, 9.3] with benign utility unchanged at 69.7\% (and 56.1\% utility under
attack, against 51.0\% undefended).

\begin{table}[h]
\centering
\caption{Built-in AgentDojo defenses (end-to-end, from the published logs) and our
  variants (trace replay) on the 76 test tasks; 95\% bootstrap intervals over
  tasks. ASR is attack success (end-to-end), and for our rows the harmful-step
  trace-level estimate. Benign utility is the share of attack-free tasks the agent
  completes; for our rows it is a proxy that keeps a task only if the undefended
  agent succeeded and no call was flagged (first value) or blocked (second value;
  flags then go to review).}
\label{tab:builtin}
\footnotesize
\setlength{\tabcolsep}{4pt}
\rowcolors{2}{gray!8}{white}
\resizebox{\linewidth}{!}{%
\begin{tabular}{@{}llccc@{}}
\toprule
\textbf{System} & \textbf{Evaluation} & \textbf{ASR [95\% CI]} &
\multicolumn{2}{c}{\textbf{Benign utility [95\% CI]}} \\
\midrule
No defense & end-to-end & 0.457 [0.376, 0.532] & \multicolumn{2}{c}{0.671 [0.566, 0.776]} \\
Spotlighting (delimiting) & end-to-end & 0.402 [0.324, 0.471] & \multicolumn{2}{c}{0.697 [0.592, 0.803]} \\
Repeat user prompt & end-to-end & 0.254 [0.189, 0.314] & \multicolumn{2}{c}{0.829 [0.737, 0.908]} \\
PI detector (classifier) & end-to-end & 0.079 [0.048, 0.114] & \multicolumn{2}{c}{0.421 [0.316, 0.526]} \\
Tool filter (LLM) & end-to-end & 0.065 [0.040, 0.093] & \multicolumn{2}{c}{0.697 [0.592, 0.803]} \\
\midrule
Tool allowlist (ours) & replay & 0.096 [0.059, 0.135] & 0.601 [0.491, 0.706] & 0.601 [0.491, 0.706] \\
Scope-only + read bypass (ours) & replay & 0.043 [0.021, 0.068] & 0.487 [0.382, 0.596] & 0.614 [0.509, 0.719] \\
Full Phase~2 (ours) & replay & 0.020 [0.007, 0.036] & 0.417 [0.316, 0.522] & 0.491 [0.390, 0.596] \\
\bottomrule
\end{tabular}}
\end{table}

The tool filter is the closest relative of our allowlist and a strong baseline:
our allowlist-only variant reaches 9.6\% harmful-step ASR [5.9, 13.5], whose
interval overlaps the tool filter's, and scope-only plus read bypass reaches 4.3\%
[2.1, 6.8], in the same range, while the full audit reaches 2.0\% [0.7, 3.6], lower
than any built-in defense. The utility comparison is less favorable to us under
our proxy (42--61\% against 69.7\%), but the two numbers are not the same kind of
measurement. Our evaluation replays logged traces and withholds a flagged call;
it therefore overstates security, because an agent might route around a block,
and understates utility, because it counts a blocked task as failed, whereas an
agent restricted to a filtered tool set may adapt, as the tool filter's
unchanged utility suggests. We do not claim that our audit beats the tool
filter. What the comparison shows is that an LLM-written allowlist alone
captures much of the benefit, that scopes and ordering buy a few more points of
attack success, and that live end-to-end runs are needed to settle their
utility cost (Section~\ref{sec:discussion} reports a pilot that could not
provide them).

\section{Results: Reasoning Guardian}
\label{sec:guardianresults}

\begin{table}[h]
\centering
\caption{Reasoning-guided escalation on the 76 test tasks (graph repetition 0,
  full Phase~2; 95\% bootstrap intervals over tasks). Only flagged calls reach
  the guardian. The first row therefore differs slightly from
  Table~\ref{tab:phase2}, which averages three graphs.}
\label{tab:guardian}
\footnotesize
\setlength{\tabcolsep}{4pt}
\rowcolors{2}{gray!8}{white}
\resizebox{\linewidth}{!}{%
\begin{tabular}{@{}lcccc@{}}
\toprule
\textbf{Configuration} & \textbf{ASR [95\% CI]} & \textbf{Benign calls flagged} &
\textbf{Benign tasks flagged [95\% CI]} & \textbf{Benign tasks blocked} \\
\midrule
Audit only                          & 0.041 [0.016, 0.073] & 0.265 [0.163, 0.375] & 0.408 [0.303, 0.526] & 0.329 \\
+ guardian (independent overrides)  & 0.041 [0.016, 0.073] & 0.182 [0.109, 0.263] & 0.368 [0.263, 0.474] & 0.289 \\
\rowcolor{tracebluelight!40}
+ guardian (state updated)          & \textbf{0.041} [0.016, 0.073] & \textbf{0.175} [0.103, 0.256] & \textbf{0.342} [0.237, 0.461] & \textbf{0.263} \\
\bottomrule
\end{tabular}}
\end{table}

\paragraph{Security is preserved.} The guardian overrode 1 of 238 flagged
malicious first calls (0.4\%, [0.1, 2.3]), and attack success is unchanged at
4.1\%. The overridden call is a \texttt{search\_emails} filtered by a sender whose
address is also consistent with the user's own task; the plan's lenient scope
allowed it, and the later send step would still be flagged. On the development
set it overrode 2 of 68 malicious calls, both of which are also ambiguous
labels (a message and a profile lookup that the user's task could require).

\paragraph{Benign flags fall.} The guardian allowed 23 of 73 flagged benign calls
(31.5\%, [22.0, 42.9]): 17 of the 56 blocks (outside the plan, or preconditions
unmet) and 6 of the 17 scope flags, mostly read-only lookups such as
\texttt{get\_webpage} and \texttt{read\_channel\_messages}. With independent
overrides, flagged benign calls fall by 8.4 points ([3.4, 14.7]) and flagged
tasks by 3.9 points ([0.0, 9.2]), which is borderline. When an override
authorizes its node, as a deployment would, flagged calls fall by 9.1 points
([3.9, 15.4]) and both flagged and blocked benign tasks by 6.6 points ([1.3, 13.2]):
from 40.8\% to 34.2\% flagged and from 32.9\% to 26.3\% blocked.

\paragraph{Cost.} The 311 decisions used 715 API calls (2.3 per decision, of
which 1.3 are expression evaluations), about 2,100 input and 90 output tokens
per decision, and a median latency of 3.6\,s (95th percentile 5.8\,s). The tool
was used in 292 decisions, with 5 evaluation errors overall. The guardian runs
on only 26.5\% of benign calls.

The gain is real but modest: about a third of benign tasks remain flagged. The
guardian cannot see tool outputs, so it cannot verify where an argument came
from, and it errs towards keeping the flag.

\paragraph{With the read-only policy.} The guardian composes with the bypass.
After the bypass it overrides 22\% of the remaining flagged benign calls in the
full design (9 of 41, [12.0, 36.7]) and 25\% in the scope-only design (7 of 28,
[12.7, 43.4]), and none of the flagged first harmful malicious calls (0 of 242
and 0 of 226; upper bounds 1.6\% and 1.7\%); all 56 new decisions kept the flag.
Harmful-step ASR is unchanged by the guardian in both designs
(Table~\ref{tab:guardianro}). Adding the guardian after the bypass lowers flagged
benign tasks by 5.3 points ([1.3, 10.5]) in the full design and 6.6 points
([1.3, 11.8]) in the scope-only design, and blocked tasks by 5.3 points
([1.3, 10.5]) and 1.3 points ([0.0, 3.9]). Relative to the original full design,
bypass plus guardian lowers flagged benign tasks from 40.8\% to 27.6\% and blocked
tasks from 32.9\% to 19.7\% (13.2 points each; [6.6, 21.1] and [5.3, 22.4]).

\begin{table}[h]
\centering
\caption{Read-only bypass and guardian (graph repetition 0, 76 test tasks, 95\%
  bootstrap intervals over tasks). ASR is measured at the harmful step. Guardian
  rows use the state-updating simulation. The scope-only design reuses the
  decisions made for the full design.}
\label{tab:guardianro}
\footnotesize
\setlength{\tabcolsep}{4pt}
\rowcolors{2}{gray!8}{white}
\resizebox{\linewidth}{!}{%
\begin{tabular}{@{}lcccc@{}}
\toprule
\textbf{Configuration} & \textbf{ASR (harmful) [95\% CI]} & \textbf{Benign calls flagged} &
\textbf{Benign tasks flagged [95\% CI]} & \textbf{Benign tasks blocked} \\
\midrule
Full design & 0.030 [0.006, 0.064] & 0.265 & 0.408 [0.303, 0.526] & 0.329 \\
\quad + read bypass & 0.030 [0.006, 0.064] & 0.149 & 0.329 [0.224, 0.434] & 0.250 \\
\rowcolor{tracebluelight!40}
\quad + guardian & 0.030 [0.006, 0.064] & 0.109 & 0.276 [0.171, 0.382] & 0.197 \\
\midrule
Scope-only design & 0.053 [0.022, 0.089] & 0.175 & 0.329 [0.224, 0.434] & 0.158 \\
\quad + read bypass & 0.053 [0.022, 0.089] & 0.102 & 0.276 [0.184, 0.382] & 0.092 \\
\rowcolor{tracebluelight!40}
\quad + guardian & 0.053 [0.022, 0.089] & 0.076 & 0.211 [0.118, 0.303] & 0.079 \\
\bottomrule
\end{tabular}}
\end{table}

\section{Results: Does Coupling Help?}
\label{sec:coupling}

\begin{table}[h]
\centering
\caption{Identity benchmark (12 principals, workspace suite, sessions of 20--30
  tasks): TPR at 5\% FPR with 95\% bootstrap intervals over sessions, using the
  scope-and-allowlist design of Phase~2. Positives: 5,946 events (splice), 3,943
  (drift), 457 (hijack). $\Delta$ is the paired difference from adding
  $\varphi_4$ to the cross-session detector.}
\label{tab:identity}
\footnotesize
\setlength{\tabcolsep}{4pt}
\rowcolors{2}{gray!8}{white}
\resizebox{\linewidth}{!}{%
\begin{tabular}{@{}lccc@{}}
\toprule
\textbf{Detector} & \textbf{Splice} & \textbf{Drift} & \textbf{Hijack} \\
\midrule
Phase~2 verdict            & 0.053 [0.047, 0.059] & 0.050 [0.043, 0.057] & 0.762 [0.718, 0.801] \\
Flag persistence (10)      & 0.058 [0.052, 0.065] & 0.058 [0.053, 0.065] & 0.120 [0.111, 0.128] \\
Online per-session $\varphi_{1\text{--}3}$   & 0.119 [0.100, 0.136] & 0.149 [0.131, 0.166] & 0.114 [0.085, 0.142] \\
Online per-session $\varphi_{1\text{--}4}$   & 0.119 [0.100, 0.136] & 0.150 [0.131, 0.167] & 0.114 [0.086, 0.141] \\
Cross-session $\varphi_{1\text{--}3}$        & 0.242 [0.199, 0.288] & 0.128 [0.088, 0.161] & 0.072 [0.037, 0.112] \\
Cross-session $\varphi_{1\text{--}4}$        & 0.195 [0.157, 0.235] & 0.091 [0.060, 0.121] & 0.121 [0.078, 0.164] \\
\rowcolor{tracebluelight!40}
$\Delta$ ($\varphi_4$, cross-session)        & $-0.047$ [$-0.060$, $-0.034$] & $-0.036$ [$-0.051$, $-0.023$] & $+0.049$ [$+0.022$, $+0.078$] \\
\bottomrule
\end{tabular}}
\end{table}

\paragraph{Splice and drift.} The attacker's requests define their own
objective, so the calls pass Phase~2 and $\gamma$ is uninformative: the
Phase~2 verdict alone is at chance (AUROC 0.48 on both). The behavioral
detectors have real but modest power: the cross-session baseline reaches TPR
0.242 on splice (AUROC 0.771) and 0.128 on drift (AUROC 0.697), and the online
per-session baseline is comparable (0.119 and 0.149; AUROC 0.733 and 0.765),
even better than the cross-session baseline on drift. Adding $\varphi_4$
\emph{lowers} the cross-session detector's detection, by 4.7 points on splice
and 3.6 points on drift, with intervals far from zero; with the full Phase~2
design the reductions are 5.6 and 4.3 points, and on the residual events
(lean design) 3.2 and 2.2 points. It makes no difference to the online
detector. In the shorter sessions (8--12 tasks, about 20 calls) the per-session
detector was active on 10\% of events and close to chance (AUROC 0.54--0.57),
and the effect of $\varphi_4$ on the cross-session detector was $-1.2$ and
$-1.1$ points ($-1.9$ and $-1.6$ with the full design): the same direction at
a smaller scale.

\paragraph{Hijack.} Here $\gamma$ does carry information, but the verdict
carries it better: Phase~2's own verdict detects 76.2\% of malicious calls at 5\%
FPR (AUROC 0.906), against 7.2\% to 12.1\% for the behavioral detectors. The
significant gain from $\varphi_4$ on hijack with the lean design (+4.9 points,
[2.2, 7.8]) only re-encodes the verdict; with the full design it is +1.8 points
([$-0.2$, 4.1]). On the residual events Phase~2 allowed, the effect is 0.0
points ([$-3.0$, 0.0]) on 59 events with the lean design and $-1.2$ points
([$-4.8$, 0.0]) on 40 events with the full design. These sets are too small to
exclude a small effect, but there is no sign of one.

\paragraph{Flag persistence.} Carrying a recent flag forward improves the
cross-session detector by 2.6 points on splice ([1.4, 4.0]) and 2.0 points on
drift ([0.5, 3.5]) with the lean design (4.5 and 3.6 points with the full
design), so even a trivial use of the verdict history adds a little where the
learned $\varphi_4$ features subtract.

\section{Discussion}\label{sec:discussion}

\paragraph{What the evidence supports.} Prospective least-privilege auditing is
effective against injection in this benchmark: the extracted plan from the
objective alone removes about 93\% of the attack success, with a deterministic
audit that costs microseconds. The price is utility: at least 15\% of benign
tasks need a tool the plan omits, and each additional check raises benign
flagging. Two inexpensive measures recover a good part of that cost. Letting
tools that only read the user's own state bypass the audit costs no
harmful-step security, because the harmful step of nearly every attack we
labelled is a write or an external call, while a large share of the benign flags
were lookups. A
reasoning guardian then removes a further share of the remaining flags without
overriding any flagged harmful call, because it is invoked only on calls the
plan has already flagged. It is the one place where a reasoning model that
executes code adds value in our results. The leanest configuration we tested,
allowlist plus argument scopes with local reads bypassed, blocks 8.8\% of benign
tasks at 4.3\% harmful-step attack success; adding the guardian flags 21.1\% of
benign tasks and blocks 7.9\% (graph 0, harmful-step ASR 5.3\%). Against the built-in defenses of AgentDojo, the LLM tool filter is the strongest
end-to-end baseline (6.5\% attack success, no utility loss), and our allowlist-only
variant is statistically indistinguishable from it on security; the case for the
additional machinery rests on the lower attack success of the scope and full
variants (4.3\% and 2.0\% at the harmful step), which our replay-based utility
proxy suggests costs benign tasks.

\paragraph{Why the coupling fails.} The alignment score is a function of a call
and the current task's graph. If the person at the keyboard changes, the new
requests re-define the task and the calls are legitimate under it, so the
score has nothing to say about who is asking. If the agent is hijacked inside a
task, the verdict itself already identifies the calls, and a behavioral model
built on top of it is a lossy copy. The idea could still work if the graph were
tied to a standing, per-principal role rather than to the current request, so
that a takeover is a deviation from the principal's established intent; we did
not test this.

\paragraph{Limitations.}
(1)~One benchmark and one agent model: AgentDojo traces of a single model, with
trace-level ASR that assumes no re-planning after a block. (2)~The extractor and
the guardian are LLMs: three graph repetitions capture sampling variance but not
variance across models or prompts, the prompt rules were tuned on four
development tasks, and the guardian was run once on a single graph repetition.
(3)~Scopes are limited by what the objective determines: 61\% of nodes leave
arguments unconstrained, and 10.6\% of malicious-call evaluations were allowed.
(4)~We did not test attacks adaptive to the audit or the guardian. Call
arguments reach the guardian and are attacker-influenced, and although we
instruct it to treat them as data, a targeted payload could try to persuade it.
(5)~The guardian cannot see tool outputs, so it cannot verify the provenance of
an argument. (6)~Some labelled malicious calls coincide in function with the
user's own task, so override rates on malicious calls may be overstated.
(7)~The read-only policy rests on a hand classification of tools and on the
assumption that harm occurs at a write or external step; an attack that leaks
data through reads alone, or through a tool we classified as a local read, would
evade it. \texttt{get\_webpage} shows that the classification matters. The
harmful-step metric was introduced after seeing that the bypass changes which
call is first, and we report both metrics. (8)~We report the lean Phase~2 design
in the identity benchmark because ordering edges raised false positives; we
chose this after seeing test-set Phase~2 results, and the full design is shown
as a robustness check with the same conclusion. (9)~The identity benchmark is
synthetic: one suite, twelve principals with overlapping task mixtures, and
modest detectors overall (TPR 0.07--0.24 at 5\% FPR), so a richer behavioral
signal may do better, although the argument above does not depend on detector
strength. Session length also matters: the per-session baseline was near chance
on sessions of about 20 calls and competitive on sessions of about 51.
(10)~We did not evaluate on production MCP logs; the internal logs available to
us had a single endpoint and shared client addresses, too little variation to
test scopes or identity. (11)~The scope compiler relies on restricted
\texttt{eval}. (12)~The comparison with built-in defenses mixes their end-to-end
evaluation with our trace replay: replay overstates security because it assumes
the agent does not route around a blocked call, and understates utility because
it counts a blocked task as failed. Live runs with the audit in the loop would
remove this gap. We built a harness for them (the audit gate inside AgentDojo's
agent loop) and ran a pilot with a current agent (gpt-5.2, benchmark v1, 10 test
tasks with 2 injection tasks each, the same attack). Without a defense the agent
followed the injection in 0 of 20 runs, although the injected text reached it in
all 20; the attack-free utility was 0.60 and the utility under attack 0.50. With
no baseline attack success there was nothing for a defense to reduce, so we did
not run the defended configurations, and the paper's security numbers remain
replay-based. The pilot also shows that the replay baseline (45.7\% attack
success) describes an older, more injection-susceptible agent, and that our
results should be read as protection against agents of that kind.

\paragraph{Reproducibility.} The analysis scripts, the cached extractor outputs and
the guardian decisions are available from the authors on request. The benchmark
logs are publicly available from the AgentDojo repository.

\section{Conclusion}

We presented an audit that turns a user's objective into an executable intent
graph with one LLM call and checks every tool call against it with deterministic
code, with a bypass for tools that only read the user's own state and a
reasoning guardian that reviews the calls that stay flagged. Under a leak-free
protocol on AgentDojo, replaying logged traces, the full audit lowers trace-level
attack success from 45.7\% to 3.3\%, at the price of flagging 40\% of benign tasks. Our ablations
show that ordering constraints are the most expensive per point of protection,
that bypassing local reads recovers utility at no cost in harmful-step security
while bypassing an external read does not, and that the guardian overrides a
quarter of the remaining flagged benign calls and none of the flagged harmful
ones. Using the audit's alignment score to improve identity-drift detection did
not improve detection beyond what the audit's own verdict already provides. The
open problems are reducing benign flags further without weakening the allowlist,
hardening the guardian against adaptive payloads, validating end to end against an injection-susceptible agent (our one live pilot, with a current agent, showed no baseline attack success),
and tying intent to a standing role so that account-level attacks become visible
to the audit.

\bibliographystyle{IEEEtran}
\bibliography{references}

\end{document}